\documentclass{pasj00}

\begin{document}
\SetRunningHead{K. Muraoka et al.}{ASTE CO(3-2) Observations of M 83}
\Received{}%{yyyy/mm/dd}
\Accepted{}%{yyyy/mm/dd}

\title{
ASTE CO(3-2) Observations of the Barred Spiral Galaxy M~83:\\ I. Correlation between CO(3-2)/CO(1-0) Ratios and\\ Star Formation Efficiencies }

%%% Please use the following style in case that sorting by 
%%% affilation is impossible. 
%
 \author{%
   Kazuyuki \textsc{Muraoka},\altaffilmark{1}
   Kotaro \textsc{Kohno},\altaffilmark{1}
   Tomoka \textsc{Tosaki},\altaffilmark{2}
   Nario \textsc{Kuno},\altaffilmark{2}\\
   Kouichiro \textsc{Nakanishi},\altaffilmark{2}
   Kazuo \textsc{Sorai},\altaffilmark{3}
   Takeshi \textsc{Okuda},\altaffilmark{1}
   Seiichi \textsc{Sakamoto},\altaffilmark{4}\\
   Akira \textsc{Endo},\altaffilmark{1,4}
   Bunyo \textsc{Hatsukade},\altaffilmark{1}
   Kazuhisa \textsc{Kamegai},\altaffilmark{1}
   Kunihiko \textsc{Tanaka},\altaffilmark{1}\\
   Juan \textsc{Cortes},\altaffilmark{4,5}
   Hajime \textsc{Ezawa},\altaffilmark{4}
   Nobuyuki \textsc{Yamaguchi},\altaffilmark{4}
   Takeshi \textsc{Sakai},\altaffilmark{2}\\
   and Ryohei \textsc{Kawabe}\altaffilmark{4}
}

 \altaffiltext{1}{Institute of Astronomy, The University of Tokyo, 
 2-21-1 Osawa, Mitaka, Tokyo 181-0015}
 \email{kmuraoka@ioa.s.u-tokyo.ac.jp}
 \altaffiltext{2}{Nobeyama Radio Observatory, Minamimaki, Minamisaku, Nagano 384-1305}
 \altaffiltext{3}{Division of Physics, Graduate School of Science, Hokkaido University, Sapporo 060-0810}
 \altaffiltext{4}{National Astronomical Observatory of Japan, 2-21-1 Osawa, Mitaka, Tokyo 181-8588}
 \altaffiltext{5}{Departmento de Astronomia, Universidad de Chile, Casilla 36-D, Santiago, Chile}

%% `\KeyWords{}' always has to be placed before `\maketitle'.
 %Do NOT move this preamble from here!

\maketitle

\begin{abstract}
We present CO($J=3-2$) emission observations with the Atacama Submillimeter Telescope Experiment (ASTE)
toward the $5' \times 5'$ (or 6.6 $\times$ 6.6 kpc at the distance $D$ = 4.5 Mpc) region of
the nearby barred spiral galaxy M~83.
We successfully resolved the major structures, i.e., the nuclear starburst region, bar, and inner spiral arms
in CO($J=3-2$) emission at a resolution of $22''$ (or 480 pc), showing a good spatial coincidence
between CO($J=3-2$) and 6 cm continuum emissions.
We found a global CO($J=3-2$) luminosity $L'_{\rm CO(3-2)}$ of $5.1 \times 10^8$ K km s$^{-1}$ pc$^2$ within the observed region.
We also found $L'_{\rm CO(3-2)}$ in the disk region ($0.5 < r < 3.5$ kpc) of $4.2 \times 10^8$ K km s$^{-1}$ pc$^2$,
indicating that CO($J=3-2$) emission in the disk region significantly contributes to the global $L'_{\rm CO(3-2)}$.
From a comparison of a CO($J=3-2$) data with CO($J=1-0$) intensities measured with Nobeyama 45-m telescope,
we found that the radial profile of CO($J=3-2$)/CO($J=1-0$) integrated intensity ratio
$R_{3-2/1-0}$ is almost unity in the central region ($r < 0.25$ kpc),
whereas it drops to a constant value, 0.6--0.7, in the disk region.
The radial profile of star formation efficiencies (SFEs), determined from 6 cm radio continuum
and CO($J=1-0$) emission, shows the same trend as that of $R_{3-2/1-0}$.
At the bar-end ($r \sim 2.4$ kpc), the amounts of molecular gas and the massive stars are enhanced
when compared with other disk regions, whereas there is no excess of $R_{3-2/1-0}$ and SFE in that region.
This means that a simple summation of the star forming regions at the bar-end and the disk cannot reproduce
the nuclear starburst of M~83, implying that the spatial variation of the dense gas fraction
traced by $R_{3-2/1-0}$ governs the spatial variation of SFE in M~83.
\end{abstract}

\section{Introduction}

In the central regions of disk galaxies, we often find intense star formation activities
(e.g., \cite{ho1997}) where not only the star formation rates (SFRs)
but also the star formation efficiencies (SFEs), defined as the  fraction of the SFR surface density
in the surface mass density of molecular gas, are also enhanced.
These star formation activities are referred to as starburst phenomena,
which are different from the star formations in the disk regions of galaxies \citep{kennicutt1998b}.
The starburst phenomena are prominent not only because they show high SFRs but also because they show elevated SFEs.
What causes these ``high SFE'' star formations?
In other words, why do SFEs differ among galaxies (e.g., \cite{young1996})
or within locations/regions in a galaxy (e.g., \cite{gao2001})?
Since stars are formed by the contraction of molecular gas, which would be initiated by
gravitational instability and/or cloud-cloud collision of molecular clouds,
it can be easily expected that some properties of molecular gas play a key role in controlling the variations of SFEs.
Investigations of molecular gas are widely performed using CO($J=1-0$) emission,
which traces the total mass of molecular gas.
In particular, extensive surveys or mappings of CO($J=1-0$) emission toward galaxies are examined 
(\cite{young1995}, \cite{kuno2000}, \cite{sorai2000a}, \cite{nishiyama2001a}, \cite{nishiyama2001b}, \cite{kuno2007}).
However, CO($J=1-0$) emission simultaneously traces both diffuse/cold molecular gas and
dense/warm molecular gas, which is directly linked to star formation.
Therefore, CO($J=1-0$) emission is useful to determine the total amount of molecular gas,
but it is difficult to compare with SFEs quantitatively.
Thus, it is essential to study {\it dense} molecular gas,
because stars are formed from the dense cores of molecular clouds.

In fact, recent studies of the dense molecular medium in galaxies based on the observations of HCN($J=1-0$) emission,
a dense molecular gas tracer ($n_{\rm H_2}$ $>$ a few $\times 10^4$ ${\rm cm}^{-3}$) due to
its large dipole moment ($\mu_{\rm HCN} = 3.0$ Debye, whereas $\mu_{\rm CO} = 0.1$ Debye),
demonstrate the intimate connection between dense molecular gas and massive star formation in galaxies.
For example, after the first report by \citet{solomon1992}, a tight correlation between HCN($J=1-0$) and 
FIR continuum luminosities has been shown among many galaxies including nearby ones
(\cite{gao2004a}, \cite{gao2004b}), and high redshift quasar host galaxies \citep{carilli2005}.
This suggests that SFEs are governed by the fraction of dense molecular gas
to the total amount of molecular contents traced by the HCN($J=1-0$)/CO($J=1-0$) intensity ratios
(e.g., \cite{solomon1992}, \cite{kohno2002}, \cite{shibatsuka2003}, \cite{gao2004a}).
A spatial coincidence between dense molecular gas traced by HCN($J=1-0$) and
massive star forming regions was also reported (\cite{kohno1999}, \cite{shibatsuka2003}).

However, HCN($J=1-0$) emission is often too weak to trace a wide-area map in the disk regions of galaxies.
HCN($J=1-0$) intensity is typically 10--20 times lower than the CO($J=1-0$) intensity in the disk regions
of the Galaxy \citep{helfer1997b} and galaxies (e.g., \cite{kohno1996}, \cite{helfer1997a}, \cite{curran2001}, \cite{sorai2002}).
Therefore, HCN($J=1-0$) emission is mainly used to measure the amount of dense molecular gas in
the central regions of galaxies or to obtain information integrated over the whole regions of galaxies
due to inadequate sensitivity of current observing facilities
(e.g. \cite{nguyen1989}, \cite{nguyen1992},  \cite{hankel1990}, \cite{solomon1992}, \cite{israel1992},
\cite{helfer1993}, \cite{aalto1995}, \cite{curran2000}, \cite{kohno2003}).

Instead, CO($J=3-2$) emission is a promising probe of dense molecular clouds even in the disk regions of galaxies.
The critical density for CO($J=3-2$) transition is a few $\times$ $10^4$ cm$^{-3}$.
Therefore, the CO($J=3-2$) emission can also trace the dense component of molecular clouds as well as HCN($J=1-0$).
In fact, in the central regions of nearby star forming galaxies, a tight correlation
between CO($J=3-2$) and H$\alpha$ luminosities, which is better than the correlation
between CO($J=1-0$) and H$\alpha$ luminosities, has been reported \citep{komugi2007}.
In addition, \citet{yao2003} discovered a trend that CO($J=3-2$)/CO($J=1-0$) ratios,
hereafter referred to as $R_{3-2/1-0}$, increase with the SFE in infrared-luminous galaxies,
which is very similar to a finding based on HCN($J=1-0$)/CO($J=1-0$) ratios (\cite{solomon1992}, \cite{gao2004a}).
Moreover, there seem to be detectable amounts of CO($J=3-2$) emitting dense gas in the disk regions
of galaxies such as M 51 (\cite{wielebinski1999}, \cite{dumke2001}), M 31 \citep{tosaki2007},
NGC 6946 \citep{walsh2002}, and Antennae galaxies \citep{zhu2003},
where the CO($J=3-2$) intensities are found to be comparable to the CO($J=1-0$) intensities.
These results clearly indicate a potential for the use of CO($J=3-2$) emission as a dense gas tracer
in the disk regions of galaxies.

Spatially-resolved wide-area CO($J=3-2$) observations of nearby galaxies are also essential
for the appropriate interpretations of high-$J$ CO observations of high redshift objects
(see a review by \cite{solomon2005}), because most of these high-$J$ CO observations of
distant objects are spatially unresolved.

Previous CO($J=3-2$) observations of nearby galaxies, however, were almost limited to their central regions
or to the whole regions of galaxies without resolving the structures using a large observing beam
(e.g. \cite{wild1992}, \cite{hurt1993}, \cite{devereux1994}, \cite{mauersberger1996}, \cite{wielebinski1999},
\cite{nieten1999}, \cite{mauersberger1999}, \cite{harrison1999}, \cite{sandqvist1999}, \cite{dumke2001},
\cite{liszt2001}, \cite{meier2001}, \cite{wiedner2002}, \cite{vila2003}, \cite{yao2003}, \cite{hafok2003},
\cite{matsushita2004}, \cite{iono2004}, \cite{narayanan2005}, \cite{weis2005}, \cite{israel2005}, \cite{israel2006}).

In this paper, we present wide-area observations of CO($J=3-2$) line emission
using the Atacama Submillimeter Telescope Experiment (ASTE; \cite{ezawa2004}, \cite{kohno2005}) towards the whole inner disk of M~83 (NGC 5236).
M~83 is a nearby, face-on, barred, grand-design spiral galaxy hosting an intense starburst at its center.
The distance to M~83 is estimated to be 4.5 Mpc \citep{thim2003};
therefore, 1$^{\prime \prime}$ corresponds to 22 pc.
The inclination of M~83 is $24^{\circ}$ \citep{comte1981}.
Its proximity and face-on view enable us to resolve its major structures such as the nuclear starburst region,
bar, and inner spiral arms even by single-dish observations at millimeter or submillimeter wavelength.
For these reasons, M~83 is the best target to investigate the spatial variations of SFE and
the physical state of the molecular gas between the nuclear region and the disk region.
The CO observations toward M~83 were reported for various transitions (see a comprehensive summary in \cite{lundgren2004}).
\citet{handa1990} present a CO($J=1-0$) map of the nucleus and the bar at a resolution of $16''$ using a Nobeyama 45-m telescope.
A new wider CO($J=1-0$) map using the Nobeyama 45-m telescope is now available \citep{kuno2007}.
Moreover, \citet{crosthwaite2002} and \citet{lundgren2004} present CO($J=1-0$) and CO($J=2-1$) maps covering the entire optical disk. 
However, the high-$J$ CO line observations such as CO($J=3-2$) were still limited to the nuclear region
and the bar (\cite{petitpas1998}, \cite{israel2001}, \cite{dumke2001}, \cite{sakamoto2004}, \cite{bayet2006}).

The goals of this paper are (1) to depict the global distribution of CO($J=3-2$) line emission
or dense molecular gas in the disk region of M~83, (2) to investigate the spatial variations of $R_{3-2/1-0}$
within the disk of M~83, by computing the radial profiles of CO($J=3-2$) and CO($J=1-0$) intensities,
and (3) to examine the correlation of the $R_{3-2/1-0}$ values and SFEs through a comparison of the radial profiles.
In particular, we focus on the difference of properties of the nuclear starburst and star formation in the disk region.
Further analysis of our CO($J=3-2$) data of M~83 will be presented in forthcoming papers.

\section{Observations and Data Reduction}

CO($J=3-2$) observations towards M~83 were performed using the ASTE
from December 2004 to January 2005, and June 2005.
The size of the CO($J=3-2$) map is about $5^{\prime} \times 5^{\prime}$ ($6.6 \times 6.6$ kpc),
including the nuclear starburst region, the bar, and the inner spiral arms.
We set the grid spacing to 11$^{\prime \prime}$, except for the inter-arm region
where the grid spacing was set to 22$^{\prime \prime}$.
We set the position angle of the major axis to 45$^{\circ}$ \citep{comte1981}.
The total number of points observed is 419, and the observed positions are indicated in figure 1.
The full-half power beam width (HPBW) is 22$^{\prime \prime}$, which corresponds to about 480 pc
at a distance of 4.5 Mpc. 

%% fig. 1
\begin{figure}
  \begin{center}
    \FigureFile(80mm,70mm){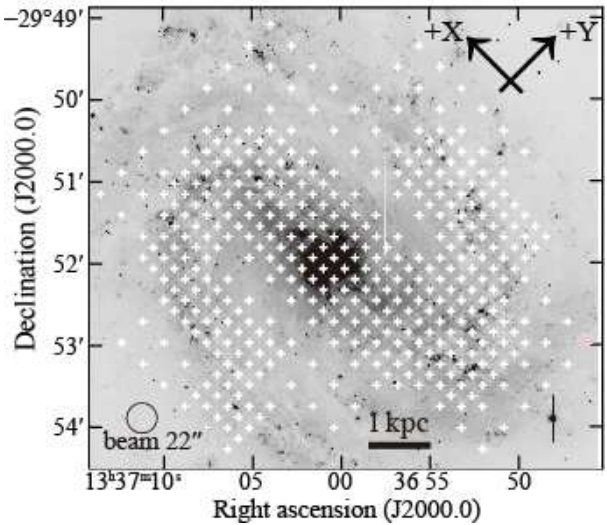}
  \end{center}
\caption{
Points observed (crosses) using the ASTE superposed on a V-band image of M~83 obtained with VLT \citep{comeron2001}.
The grid points are spaced by 11$^{\prime \prime}$ at the center, bar, and along the spiral arms
and by 22$^{\prime \prime}$ in the inter-arm.
}\label{fig:obsp}
\end{figure}

We used a cartridge-type cooled SIS mixer receiver, SC345, developed by collaboration
among the University of Tokyo, the Osaka Pref.\,University, the Nagoya University,
and the National Astronomical Observatory Japan (NAOJ).
The receiver SC345 was plugged into the Single Cartridge Dewar of the ASTE, which is based on
the design of the ALMA test cryostat (\cite{sekimoto2003}, \cite{sugimoto2003}, \cite{yokogawa2003})
using a 3-stage Gifford McMahon cryocooler with an He pot for temperature stabilization.
A Nb-based parallel-connected twin-junction-type mixer is mounted on the 4 K stage of the dewar
with a corrugated feed horn and a local oscillator (LO) signal coupler to inject radio-frequency (RF)
and LO signals into the receiver. The intermediate-frequency (IF) range is 4.5 to 7.0 GHz.
An LO system is composed of a chain of multipliers, i.e., an active frequency doubler,
two frequency doublers, and a frequency tripler, provided by Virginia Diode Inc.
A source signal generated by a GPS-locked frequency synthesizer with a frequency range
of 13.5 GHz to 15.5 GHz is multiplied by 24 using the multiplier chain system.
Both the attenuation of the LO signal power and the change in the frequency setting without a phase lock loop of Gunn oscillator
easily facilitate fully remote operations of the receiver system. 
The resultant LO frequency range is 324 to 372 GHz, and the typical receiver noise temperature is about 100 K
(in double side-band) for this frequency range.
The average system temperature was 200 K (in double side-band) during our observations of M~83.
The temperature scale was calibrated by the chopper-wheel method,
using a room temperature single absorber in front of the receiver.
The observations were made remotely from the ASTE operation rooms of
the Institute of Astronomy (IoA), Univ. of Tokyo, NAOJ, and the Nobeyama Radio Observatory (NRO)
using a network observation system N-COSMOS3 developed at NAOJ (\cite{kam05}).
The sky emission was subtracted by position switching with two off-source positions 
at offsets in an azimuth of $\pm$ 10$^{\prime}$ from the central position, which was taken
at $\alpha = 13^h 37^m 00^s .48$, $\delta = -29^{\circ} 51^{\prime} 56^{\prime \prime}.5$ (J2000).
This position corresponds to the radio continuum center \citep{sukumar1987}.
The spectra of the CO($J=3-2$) emission were obtained with four units of XF-type digital spectrometers,
MAC \citep{sorai2000b}, with a band width of 512 MHz and 1024 channels,
corresponding to a velocity coverage of 440 km s$^{-1}$ for each unit.
The absolute pointing of the antenna was checked every hour using Jupiter and 
the accuracy was found to be better than $\sim 2^{\prime \prime}$ rms.

We observed the CO($J=3-2$) emission of IRC+10216 twice a day to verify the stability of the main beam efficiency ($\eta_{\rm MB}$)
of the ASTE 10-m dish, and we have compared our data with the CO($J=3-2$) emission of IRC+10216 obtained by
CSO observations \citep{wang1994}, which are carried out using an SSB-filter.
For the CSO observations, the line peak temperature of CO($J=3-2$) emission of IRC+10216 in $T_{\rm MB}$ was 32.5 $\pm$ 0.5 K.
Our CO($J=3-2$) data was calibrated using this value.
The resultant $\eta_{\rm MB}$ obtained by the calibration of the line temperature was 0.63 $\pm$ 0.06.
In addition to this, we have also calculated $\eta_{\rm MB}$ using the brightness temperature of Jupiter
recorded during the pointing observation.
We assumed that the brightness temperature of Jupiter at 345 GHz is 174 K \citep{mangum1993}.
The resultant $\eta_{\rm MB}$ deduced from the Jupiter measurements was 0.62 $\pm$ 0.03.
These $\eta_{\rm MB}$ values obtained by the two methods are consistent,
indicating that the side-band gain ratio of the receiver SC345 was almost 1.

Data reduction were carried out using NEWSTAR and GBASE, tools for single-dish data analysis
developed by NAOJ. The linear baselines were removed from most of the spectra. 
We binned the adjacent channels to a velocity resolution of 5 km s$^{-1}$ for the CO($J=3-2$) spectra.
The resultant rms noise level was typically in the range of 40 to 70 mK in the $T_{\rm MB}$ scale.
The observation parameters for M~83 are listed in table 1.
Also, the adopted parameters of M~83 are listed in table 2.

%%%%%%%%%%%
% Table 1

\begin{table}
\begin{center}
Table~1.\hspace{4pt}Observation parameters of M~83.\\[1mm]
\begin{tabular}{ll}
\hline \hline \\[-3mm]
Telescope, Receiver & ASTE 10-m, SC345\\
Spectrometer & 512 MHz, 1024 channel\\
Velocity coverage & 440 ${\rm km}$ ${\rm s}^{-1}$\\
Map description & 419 positions, $5^{\prime} \times 5^{\prime}$\\
rms noise level & 40--70 mK (in $T_{\rm MB}$ scale)\\
Beam size & $22^{\prime \prime}$\\
Main-beam efficiency$^{\ast}$ & 0.62 $\pm$ 0.06 $\pm$ 0.03\\
\hline \\[-2mm]
\end{tabular}\\
{\footnotesize
$^{\ast}\hspace{4pt}$For the main-beam efficiency, the first error indicates\\[-0.5mm]
systematic error and the second, random error.
}
\end{center}
\end{table}

%%%%%%%%%%%
% Table 2

\begin{table}
\begin{center}
Table~2.\hspace{4pt}Adopted parameters of M~83.\\[1mm]
\begin{tabular}{ll}
\hline \hline \\[-3mm]
Map center$^a$: & \\
\,\,\,R.A. (J2000) & $13^{\rm h} 37^{\rm m} 00^{\rm s}.48$\\
\,\,\,Decl. (J2000) & $-29^{\circ} 51^{\prime} 56^{\prime \prime}.5$\\
Distance$^b$ & 4.5 Mpc\\
Linear scale & 22 pc arcsec$^{-1}$\\
Systemic velocity (LSR)$^c$ & 510 km s$^{-1}$\\
Position angle$^d$ & 45$^{\circ}$\\
Inclination$^d$ & 24$^{\circ}$\\
\hline \\[-2mm]
\end{tabular}\\
{\footnotesize
Reference\,---. $^a$\citet{sukumar1987}, $^b$\citet{thim2003},\\[-0.5mm]
$^c$this paper, $^d$\citet{comte1981}
}
\end{center}
\end{table}

\section{Results}

\subsection{CO($J=3-2$) Spectra}

Figure 2 shows all 419 spectra of the CO($J=3-2$) emission of M~83.
The X and Y axes were set parallel to the major and minor axes of the galaxy, respectively.
In addition, figure 3 shows the spectra in the central $44^{\prime \prime} \times 44^{\prime \prime}$ region.
Significant CO($J=3-2$) emission for which the signal to noise ratio exceeds 4 is detected
for 275 points out of the 419 observed points.
Strong emission is observed at the nucleus and the leading edges of the bar.
Moreover, the emission line is significantly detected from most of the points along the inner spiral arms.
No significant emission lines are detected in the inter-arm regions.

%% fig. 2
\begin{figure*}
  \begin{center}
    \FigureFile(170mm,155mm){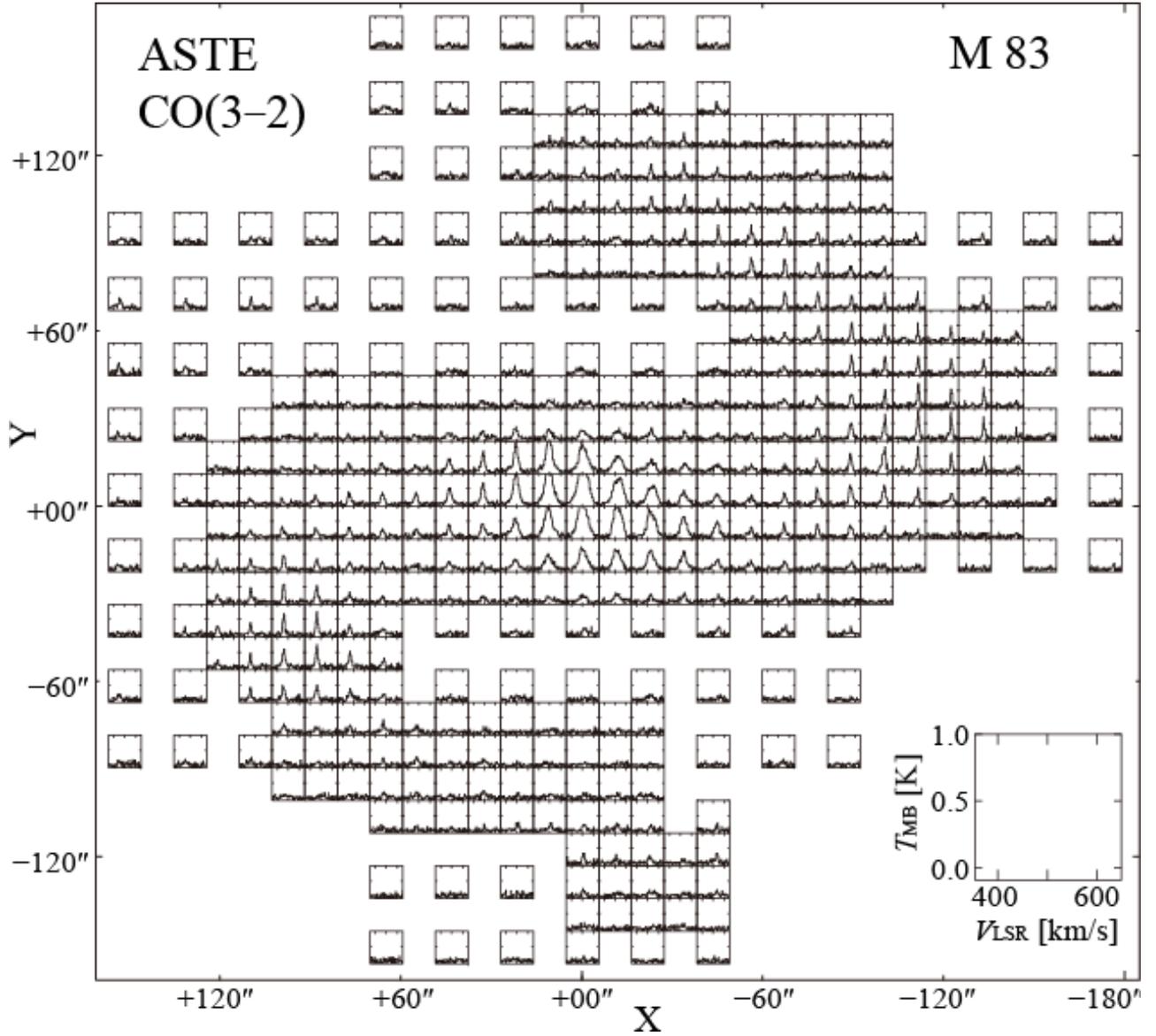}
  \end{center}
\caption{
All measured spectra of CO($J=3-2$) emission of M~83.
The scale of the spectra is indicated by the small box inserted in the lower right corner.
Significant emission is detected not only at the nucleus but also at the bar and inner spiral arms.
The central $5 \times 5$ spectra are magnified and shown in figure 3.
}\label{fig:allspec}
\end{figure*}

The physical properties of the observed CO($J=3-2$) emission line of M~83 are summarized in table 3.
Velocity-integrated intensities, $I_{\rm CO(3-2)}$, were calculated from the CO emission
within a velocity range of 380 to 620 km s$^{-1}$ of spectra.
In addition, we calculated the luminosity of CO($J=3-2$), $L^{\prime}_{\rm CO(3-2)}$.
We found that $L^{\prime}_{\rm CO(3-2)}$ was $9.3 \times 10^7$ K km s$^{-1}$ pc$^2$ in the central region ($r$ $<$ 0.5 kpc)
and $4.2 \times 10^8$ K km s$^{-1}$ pc$^2$ in the disk region (0.5 $<$ $r$ $<$ 3.5 kpc), respectively.
The resultant global CO luminosity was found to be $5.1 \times 10^{8}$ K km s$^{-1}$ pc$^2$.
This means that $L^{\prime}_{\rm CO(3-2)}$ in the disk region significantly contributes
to the global $L^{\prime}_{\rm CO(3-2)}$.
Note that the global $L^{\prime}_{\rm CO(3-2)}$ is still the lower limit
because we did not observe CO($J=3-2$) emission toward the outer disk region ($r$ $>$ 3.5 kpc).

%%%%%%%%%%%
% Table 3

\begin{table*}
\begin{center}
Table~3.\hspace{4pt}Physical properties of observed CO($J=3-2$) emission line of M~83.\\[1mm]
\begin{tabular}{ccccc}
\hline \hline \\[-3mm]
\multicolumn{2}{c}{Region} & $I_{\rm CO(3-2)}$ & $L^{\prime}_{\rm CO(3-2)}$ & $R_{3-2/1-0}$ \\
 & & [K km s$^{-1}$] & [K km s$^{-1}$ pc$^2$] & \\
\hline \\[-3mm]
M~83 & center ($r$ $<$ 0.5 kpc) & 120 & 9.3 $\times$ 10$^7$ & 0.97 $\pm$ 0.10 \\
    & disk (0.5 $<$ $r$ $<$ 3.5 kpc)  & 10 & 4.2 $\times$ 10$^8$ & 0.65 $\pm$ 0.07 \\
    & global & 12 & 5.1 $\times$ 10$^8$ & 0.69 $\pm$ 0.07 \\
\hline \\[-2mm]
\end{tabular}\\

\end{center}
\end{table*}

Now, we compare $L^{\prime}_{\rm CO(3-2)}$ of M~83 to those of other galaxies.
\citet{yao2003} gave the plot of $L^{\prime}_{\rm CO(3-2)}$ vs. $L_{\rm FIR}$
for 54 galaxies of which $L_{\rm FIR}$ is in the range of $10^9 \LO$ to $10^{12} \LO$. 
$L_{\rm FIR}$ of M~83 is $\sim 9 \times 10^9 \LO$ \citep{mauersberger1999},
and the average $L^{\prime}_{\rm CO(3-2)}$ in Yao's sample is $\sim 10^{4.5}$ K km s$^{-1}$ Mpc$^2$ $\Omega_b$
at $L_{\rm FIR} = 9 \times 10^9 \LO$. Because $10^{4.5}$ K km s$^{-1}$ Mpc$^2$ $\Omega_b$ in \citet{yao2003}
corresponds to $2 \times 10^8$ K km s$^{-1}$ pc$^2$, $L^{\prime}_{\rm CO(3-2)}$ of M~83
determined by our wide-area CO($J=3-2$) observations is comparable to that of Yao's sample.

The peak temperature and the integrated intensity at the nucleus ($r$ $<$ 250 pc) are
1.51 $\pm$ 0.04 K and 161 $\pm$ 4 K km s$^{-1}$, respectively, in the $T_{\rm MB}$ scale.
This integrated central CO value is in excellent agreement with 167 $\pm$ 15 K km s$^{-1}$
for 21$^{\prime \prime}$ beam reported by \citet{israel2001}.
Note that \citet{mauersberger1999} report the CO($J=3-2$) integrated intensity at the nucleus of M~83
of 126 K km s$^{-1}$, whereas \citet{dumke2001} found that the line intensity was 234 $\pm$ 14 K km s$^{-1}$,
although they used the same 10-m telescope, the Heinrich-Hertz-Telescope. 
Our measured CO($J=3-2$) integrated intensity lies between these two values.
It is unclear what causes such a significant discrepancy. 
There is slight difference in the position of the ``center'', a few arcsec,
yet it is clearly insufficient to account for the large discrepancy.

At the bar-end (110$^{\prime \prime}$ or 2.4 kpc away from the center to the X direction of M~83; \cite{kuno2007}), 
where secondary intensity peaks exist (see also figure 4 and figure 7a), 
the peak temperatures are in the range of 0.5 K to 0.9 K.
The typical line width is about 50 km s$^{-1}$.
In the spiral arm region, the line temperature is in the range of 0.3 to 0.5 K.
In the spectra of M 51 (\cite{wielebinski1999}), the peak temperatures of the CO($J=3-2$) emission
in the spiral arms are 0.3--0.4 K, which are comparable to those of M~83.

\subsection{Global Distributions of the CO($J=3-2$) Emission}

Figure 4 shows an integrated intensity map and a peak temperature map of the CO($J=3-2$) emission of M~83.
Extended CO($J=3-2$) emission along the bar can be clearly seen as well as a prominent concentration of CO($J=3-2$)
emission toward the nucleus. The extended components are in the leading sides of the bar;
they are often referred to as ``molecular offset ridges,''
which are visible in high resolution CO($J=1-0$) maps of barred galaxies.
A strong CO($J=3-2$) emission is also evident along the spiral arms;
in particular, we can clearly trace two ``CO($J=3-2$) arms'' in the peak temperature map.
Note that they are not so clear in the integrated intensity map due to
the narrow line width and weaker intensity of CO in the spiral arms.

Previous CO($J=3-2$) observations of M~83 were confined to a small portion ($\sim 1' - 2'$) of the galaxy
(\cite{petitpas1998}, \cite{israel2001}). \citet{dumke2001} mapped a wider region ($\sim 2' \times 2'$);
however, detected CO($J=3-2$) emission remained confined within a region along the bar.
Thus, this is the first map that clearly depicts the distribution
of warm and dense molecular gas clouds in the disk region of M~83 including spiral arms.
Basically, the overall appearance of our CO($J=3-2$) maps is similar to those of
previous CO($J=1-0$) and CO($J=2-1$) maps presented by \citet{handa1990},
\citet{crosthwaite2002}, \citet{lundgren2004}, \citet{kuno2007}.
Our high-quality data set allows us to see the molecular gas
structures more clearly through CO($J=3-2$) emission.

%% fig. 3
\begin{figure}
  \begin{center}
    \FigureFile(80mm,73mm){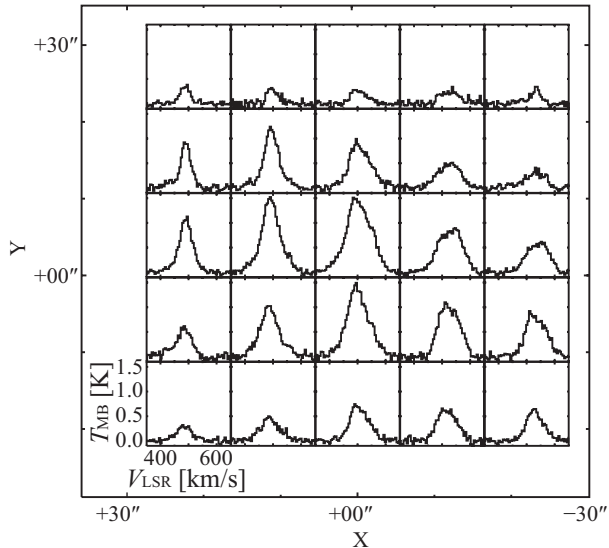}
  \end{center}
\caption{
The central $5 \times 5$ spectra of M~83. Grid spacing is 11$''$.
}\label{fig:centspec}
\end{figure}

%% fig. 4
\begin{figure*}
  \begin{center}
    \FigureFile(170mm,88mm){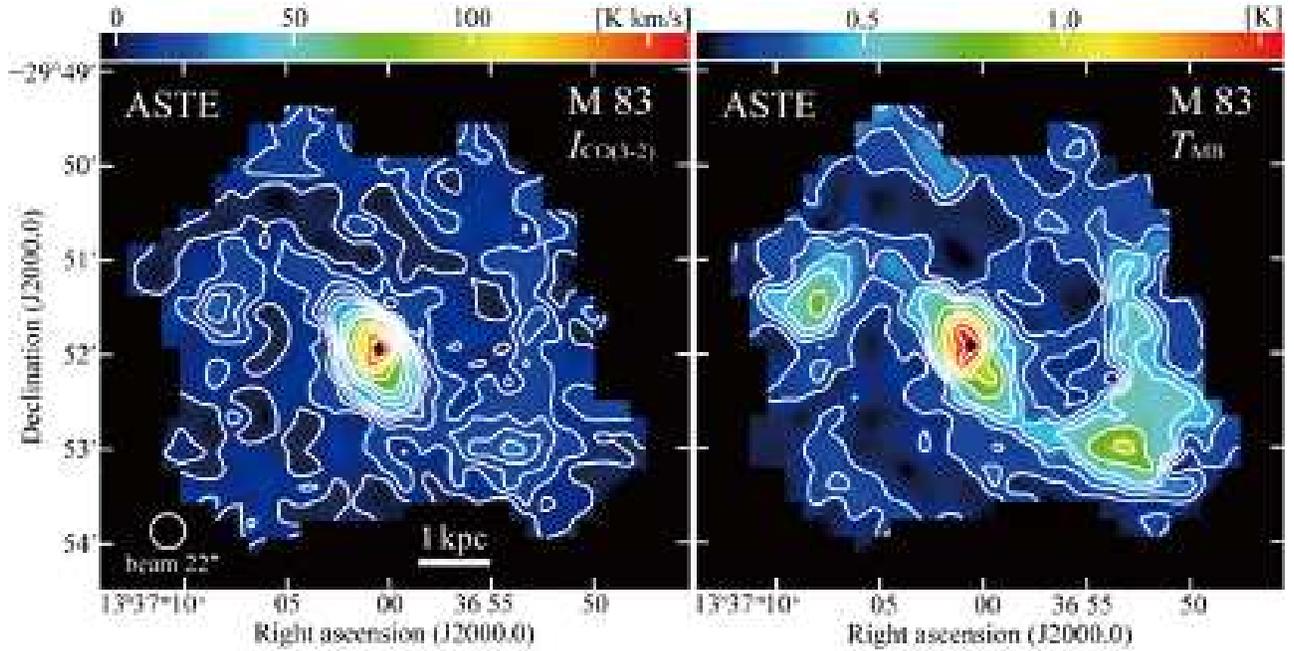}
  \end{center}
\caption{
(left)
A contour map of the CO($J=3-2$) integrated intensity of M~83. The central cross indicates
the central reference position of the map.
Contours are at 5, 10, 15, 20, 25, 30, 40, 50, 60, 80, 100, 120, 140, and 160 ${\rm K\,km\,s^{-1}}$,
and the peak is 161.2 ${\rm K\,km\,s^{-1}}$. The 1 $\sigma$ noise level of the map is 2 ${\rm K\,km\,s^{-1}}$.
(right)
A contour map of the CO($J=3-2$) peak brightness temperature. Contours are at 0.2, 0.3, 0.4, 0.5, 0.7, 0.9, 1.1, 1.3, and 1.5 K,
and the peak is 1.53 K.
Major structures, i.e., the nuclear starburst region, bar, and inner spiral arms are clearly resolved.
}\label{fig:maps}
\end{figure*}

In figure 5, we see a striking similarity between the contour map of the CO($J=3-2$) peak temperature 
and a 6 cm radio continuum map obtained with the VLA (\cite{ondrechen1985}, \cite{neininger1993}),
which is convolved to 22$^{\prime \prime}$ resolution to compare it with our CO($J=3-2$) image.
The centimeter continuum emission is mainly dominated by nonthermal synchrotron radiation
from supernova remnants in M~83 (\cite{ondrechen1985}, \cite{turner1994}).
The observed radio intensity is proportional to a recent star formation rate.
We find that these maps are well correlated, indicating a good spatial correlation
between star formation and the dense and warm gas traced by the CO($J=3-2$) emission.

Then, we examine a comparison of CO($J=3-2$) intensity with CO($J=1-0$) intensity.
We use new CO($J=1-0$) data obtained using the Nobeyama 45-m telescope (\cite{kuno2007}).
The HPBW of the CO($J=1-0$) observations with the 45-m telescope is $\sim 16^{\prime \prime}$,
and the central CO($J=1-0$) intensity is 170 K km s$^{-1}$ in the $T_{\rm MB}$ scale.
Here, we compare the new CO($J=1-0$) data with previous CO($J=1-0$) data presented by \citet{handa1990}.
According to \citet{handa1990}, the central CO($J=1-0$) intensity at a resolution of
16$^{\prime \prime}$ was $\sim 80$ K km s$^{-1}$ in the $T_{\rm A}^{\ast}$ scale.
Since the beam efficiency of their observations was 0.45, the resultant central CO($J=1-0$)
intensity in the $T_{\rm MB}$ scale was 180 K km s$^{-1}$. 
Therefore, the new CO($J=1-0$) data is in excellent agreement with the previous CO($J=1-0$) data.

Figure 6 displays a plot of CO($J=1-0$) integrated intensity vs. CO($J=3-2$) integrated intensity
for each observed point. To match the beam size of CO($J=1-0$) to that of CO($J=3-2$),
we convolved the CO($J=1-0$) data of 16$^{\prime \prime}$ resolution to 22$^{\prime \prime}$ resolution.
The red circles indicate the observed points in the central region ($r$ $<$ 0.5 kpc) of M~83
and the black ones, the inner disk region (0.5 $<$ $r$ $<$ 3.5 kpc).
This plot excludes the data points where the CO($J=1-0$) intensities are lower than 8 K km s$^{-1}$,
corresponding to a typical 3 $\sigma$ rms.
At the center of M~83, the CO($J=3-2$) intensity is close to or greater than the CO($J=1-0$) intensity,
whereas the CO($J=3-2$) intensities are weaker than the CO($J=1-0$) intensity
for most of the data points in the disk region.

The azimuthally averaged radial profiles of CO($J=3-2$) and CO($J=1-0$) integrated intensities are shown in figure 7a.
The radial profile of CO($J=3-2$) is similar to that of CO($J=1-0$);
both of them have strong peaks at the nucleus and a secondary peak at the bar-end ($\sim 2.4$ kpc),
where the molecular gas is accumulated. Such a radial CO profile is often observed
in barred spiral galaxies \citep{nishiyama2001b}.
Note that the radial profile of CO($J=3-2$) integrated intensities for larger radii (1 $<$ $r$ $<$ 3.5 kpc)
are biased toward the intensites in the bar and the inner spiral arms 
since the observed points of the CO($J=3-2$) emission are biased toward the bar and the spiral arms.
The same bias exists in the radial profile of $R_{3-2/1-0}$,
which will be described in the following subsection.

\subsection{CO($J=3-2$)/CO($J=1-0$) Intensity Ratio $R_{3-2/1-0}$}

We have obtained $R_{3-2/1-0}$ for each observed point.
The absolute error of the CO($J=3-2$) and CO($J=1-0$) intensities are estimated to about 15 \%, respectively.
Thus, the absolute error of $R_{3-2/1-0}$ is estimated to be about 30\%.

The central $R_{3-2/1-0}$ at a resolution of 22$^{\prime \prime}$ is 1.1.
For the center of M~83, \citet{dumke2001} reported $R_{3-2/1-0}$ at a resolution
of 22$^{\prime \prime}$ is $\sim 1.4$. This discrepancy is mainly due to
their high CO($J=3-2$) integrated intensity value than our CO($J=3-2$) data (see section 3.1).

Note that \citet{kramer2005} reported that the central $R_{3-2/1-0}$ is 0.6
at a resolution of 80$^{\prime \prime}$. Unfortunately, we cannot convolved our CO($J=3-2$) data
to 80$^{\prime \prime}$ due to imperfect sampling in the disk region.
Therefore, we cannot compare $R_{3-2/1-0}$ at a resolution of 80$^{\prime \prime}$
with that of \citet{kramer2005}.

A radial profile of $R_{3-2/1-0}$ is displayed in figure 7b.
$R_{3-2/1-0}$ is almost unity in the nuclear region ($r$ $<$ 0.25 kpc),
whereas it drops to a constant value in the regions with larger radii (1 $<$ $r$ $<$ 3.5 kpc).
The mean value of the ratio is 0.65 $\pm$ 0.07 within the inner disk region.

%% fig. 5
\begin{figure}
  \begin{center}
    \FigureFile(80mm,80.4mm){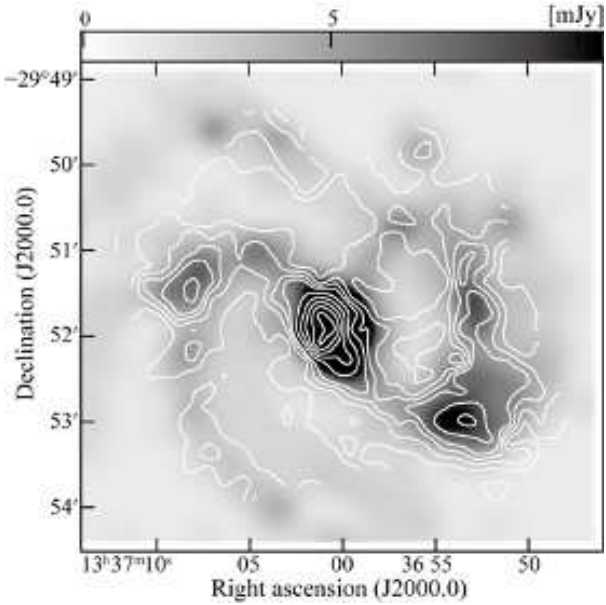}
  \end{center}
\caption{
The CO($J=3-2$) peak temperature map (white contour) superposed on the 6 cm radio continuum map (grey scale) obtained with the VLA \citep{ondrechen1985}.
These two maps correlate well spatially.}
\label{fig:3210plot}
\end{figure}

%% fig. 6
\begin{figure}
  \begin{center}
    \FigureFile(80mm,77mm){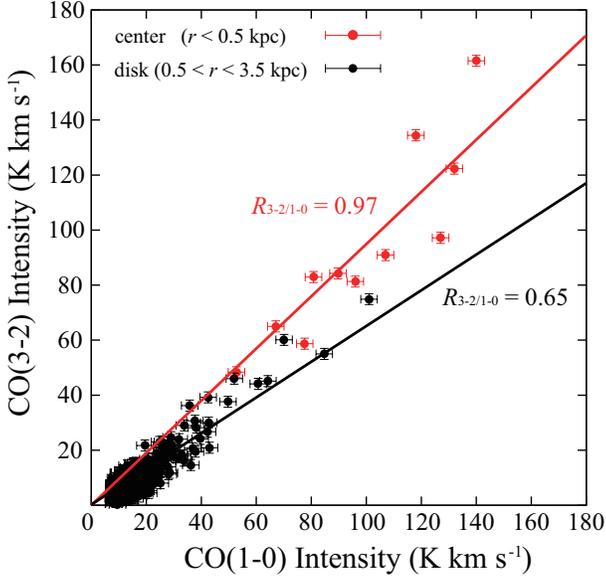}
  \end{center}
\caption{
A plot of the CO($J=1-0$) intensity vs. CO($J=3-2$) intensity for each observed point of M~83.
Data points where CO($J=1-0$) intensities are lower than 8 K km s$^{-1}$ are excluded.
The red circles indicate the points in the central region ($r$ $<$ 0.5 kpc)
and the black ones, the inner disk region (0.5 $<$ $r$ $<$ 3.5 kpc).
In this plot, the error originated from the noise is only included,
but errors from uncertainties of baselines and calibrations are not included.
The red line represents the average $R_{3-2/1-0}$ in the central region, 0.97,
while the black line represents that in the disk region, 0.65.
At the center of M~83, the CO($J=3-2$) intensities are close to or greater than CO($J=1-0$),
whereas the CO($J=3-2$) intensities are weaker than CO($J=1-0$) for most of the data points in the disk region.
}\label{fig:tmb6cm}
\end{figure}

%% fig. 7
\begin{figure}
  \begin{center}
    \FigureFile(80mm,111mm){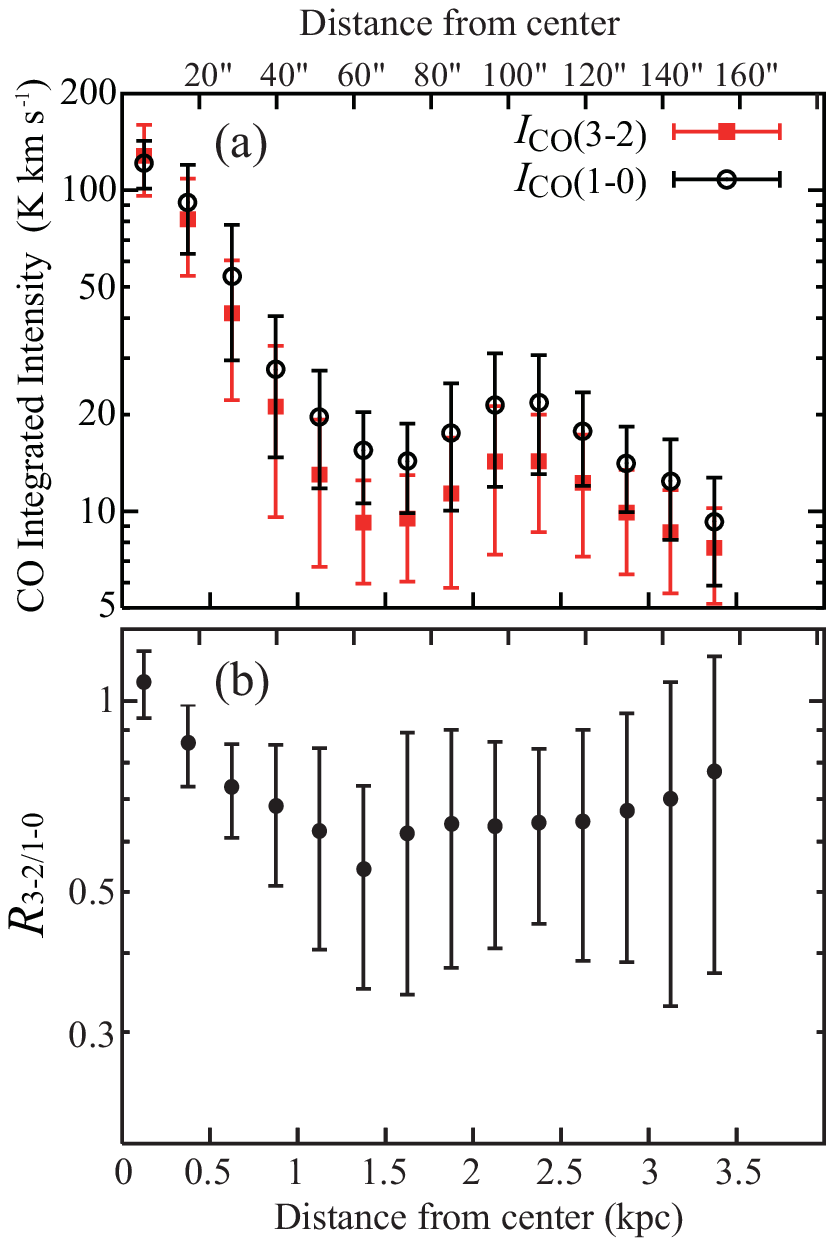}
  \end{center}
\caption{
(a)
CO($J=3-2$) and CO($J=1-0$) line intensities as a function of the galactocentric radius of M~83.
Both profiles show strong peaks in the central region as well as secondary peaks at the bar-end ($r \sim 2.4$ kpc).
(b)
$R_{3-2/1-0}$ as a function of the galactocentric radius of M~83.
$R_{3-2/1-0}$ is almost unity at the nucleus ($r$ $<$ 0.25 kpc) and drops to
a constant value, 0.6--0.7, in the disk region (0.5 $<$ $r$ $<$ 3.5 kpc),
showing no secondary peak at the bar-end.
}\label{fig:icoratio}
\end{figure}

There is a discrepancy between this radial profile of the $R_{3-2/1-0}$ and the CO($J=2-1$)/CO($J=1-0$)
integrated intensity ratio $R_{2-1/1-0}$ presented by \citet{lundgren2004}.
Their $R_{2-1/1-0}$ radial profile has a peak not at the nucleus
but at the radius of 1 kpc, corresponding to the bar region.
The reason for this discrepancy is unclear, although the difference
in the observing beams (our $R_{3-2/1-0}$ is obtained at a $22^{\prime \prime}$ resolution,
whereas their $R_{2-1/1-0}$ is at a $49^{\prime \prime}$ resolution) may partly account for this.

In other galaxies, the trend of the spatial variations of $R_{2-1/1-0}$ is similar to 
that of $R_{3-2/1-0}$ in M~83.
For example, the reported $R_{2-1/1-0}$ in the central 900 pc of the Galaxy is 0.96 (\cite{oka1998}, \cite{sawada2001}),
which is higher than the typical value of $R_{2-1/1-0}$ in the Galactic disk, 0.6--0.7 (\cite{sakamoto1995}, \cite{sakamoto1997}).
Further, for the Large Magellanic Cloud, \citet{sorai2001} reported that
there existed a radial gradient of $R_{2-1/1-0}$ of 0.94 in the inner region
($<$2 kpc from the kinematic center) and 0.69 in the outer region.

We divided the $R_{3-2/1-0}$ values of M~83 into two groups, i.e.,
the central region ($r$ $<$ 0.5 kpc) and the disk region (0.5 $<$ $r$ $<$ 3.5 kpc).
The intensity-weighted (summation of the intensity in each $R_{3-2/1-0}$ range) histograms of
$R_{3-2/1-0}$ for these two groups are shown in figure 8.
These histograms include only the data for which signal to noise ratio of the CO($J=3-2$) emission exceeds 5.
The number of observed points used to make this histogram is 11 for the center and 219 for the disk.

%% fig. 8
\begin{figure}
  \begin{center}
    \FigureFile(80mm,116mm){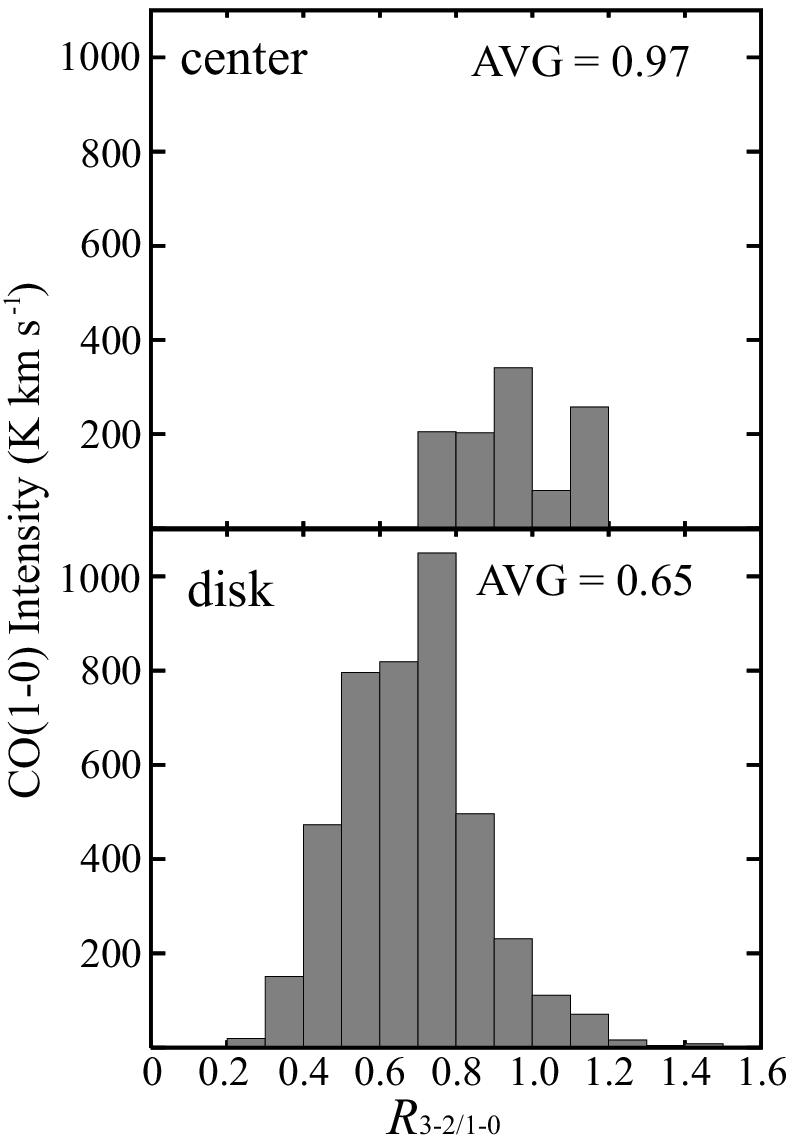}
  \end{center}
\caption{
Intensity-weighted histograms of $R_{3-2/1-0}$ of M~83.
(top)
at the center ($r$ $<$ 0.5 kpc).
(bottom)
in the disk region (0.5 $<$ $r$ $<$ 3.5 kpc).
}\label{fig:hist}
\end{figure}

In the central region ($r$ $<$ 0.5 kpc) of M~83, $R_{3-2/1-0}$ is in the range of 0.7 to 1.2,
and the average value is 0.97.
This is comparable with the average value of $R_{3-2/1-0}$ in nearby starburst galaxies,
$\sim 0.9$ (\cite{devereux1994}), and slightly higher than the average values of $R_{3-2/1-0}$
in compact galaxies, $\sim 0.7$ \citep{israel2005};
infrared luminous galaxies, $\sim 0.6$ \citep{yao2003};
dwarf starburst galaxies, $\sim 0.6$ (\cite{meier2001}); and early type galaxies, $\sim 0.4$
(\cite{vila2003}). This could be partly due to the difference in the aperture sizes
where the $R_{3-2/1-0}$ values were measured (e.g., \cite{meier2001}).
Moreover, our $R_{3-2/1-0}$ in the central region of M~83, 0.97, appears to be slightly higher than
the average values of $R_{3-2/1-0}$ in nearby galaxies, $\sim 0.6$ \citep{mauersberger1999}.
However, as we pointed out in section 3.1, $I_{\rm CO(3-2)}$ at the nucleus of M~83
reported by \citet{mauersberger1999} is about 30\% lower than our result.
Therefore, we cannot simply compare these two average values.
The reported $R_{2-1/1-0}$ are also close to unity ($\sim 0.7$ to 1; \cite{crosthwaite2002}, \cite{lundgren2004}),
although their aperture sizes are larger than ours by a factor of 2 or more.

On the other hand, the ratio in the inner disk spreads over a wide range
of 0.2 to 1.4, and the average value is 0.65.
More than half of the data points in the disk region show $R_{3-2/1-0}$ values ranging from 0.5 to 0.8.
These $R_{3-2/1-0}$ values are slightly higher than the average value of cold giant molecular clouds (GMCs)
in the Galactic disk, 0.4 \citep{sanders1993}. 
Note that there exist four data points for which $R_{3-2/1-0}$ exceeds 1.2 even in the disk region.
These four points are in the spiral arm regions but are not adjacent to each other, i.e. existing locally.
A similar situation has been observed in a spiral arm of the Milky Way galaxy.
$R_{3-2/1-0}$ in the Sagittarius arm of the Galaxy is $\sim 0.5$,
but there exist locations at which $R_{3-2/1-0}$ exceeds unity (T., Sawada. private communication).

\section{Discussion}

Thanks to our high-quality data set of CO($J=3-2$) emission, 
we have obtained the distribution of wide-ranging $R_{3-2/1-0}$,
which can resolve the 500 pc scale structure.
As mentioned in section 3, $R_{3-2/1-0}$ is almost unity at the nucleus,
and then drops to a constant value (0.6--0.7) in the disk regions.

In this section, we discuss the relation between the physical state of molecular gas
and star-formation activity using the data on $R_{3-2/1-0}$.

\subsection{Interpretation of $R_{3-2/1-0}$}

To obtain the physical interpretation of the observed $R_{3-2/1-0}$ values,
we have examined the dependence of the $R_{3-2/1-0}$ values on the kinetic temperature and gas density
by employing the large velocity gradient (LVG) approximation (\cite{scoville1974}, \cite{goldreich1974}).
Figure 9a shows the result of the LVG calculations, assuming Z($^{12}$CO)/$dv/dr = 8 \times 10^{-5}$,
which would be a typical value for galaxies.
If we consider the one-zone model to be valid and assume the kinetic temperature $T_{\rm kin} \sim 20$ K,
which seems to be relevant for GMCs in the disk regions,
it appears that $R_{3-2/1-0}$ depends mainly on the density of gas rather than
the kinetic temperature in the disk region where the $R_{3-2/1-0}$ is in the range of 0.1 to 0.7.
The ${\rm H_2}$ gas density is $\sim 10^{2.5-3}$ ${\rm cm}^{-3}$ in the disk region.
However, in this assumption of Z($^{12}$CO)/$dv/dr$ = $8 \times 10^{-5}$, we cannot
reproduce the high $R_{3-2/1-0}$ values observed in the central region,
where $R_{3-2/1-0}$ is close to or exceeds unity, suggesting a low opacity of CO emission
or difference in emitting regions of a clump between CO($J=3-2$) and CO($J=1-0$) due to external heating \citep{gierens1992}.
Therefore, in the central region, we must assume another condition different from that in the disk region.
Figure 9b shows the result of the LVG calculations, assuming Z($^{12}$CO)/$dv/dr = 8 \times 10^{-6}$.
In this condition, the kinetic temperature is higher than 60 K and the ${\rm H_2}$ gas density is
$\sim 1 \times 10^4$ ${\rm cm}^{-3}$ at the nucleus where the $R_{3-2/1-0}$ exceeds unity.
This is a fairly strong constraint, and the molecular gas is in a significantly warm and dense state.
From these results of the LVG approximation, the variation of $R_{3-2/1-0}$ in the range
of 0.4 to 1.0 seems to depend mainly on the density of the molecular gas.
Note that in the condition where $R_{3-2/1-0}$ is close to unity,
$R_{3-2/1-0}$ also restricts the kinetic temperature as well as the density of the molecular gas.

In the case of external galaxies, the observing beams are usually
too large to distinguish the individual molecular cloud structures,
and the one-zone assumption is no longer valid.
Thus, $R_{3-2/1-0}$ could be a rough measure of the fraction of the dense molecular gas ($n_{\rm H_2} >$ a few $\times 10^3$ ${\rm cm}^{-3}$)
to the total (including diffuse) molecular clouds within the observing beam.

%% fig. 9
\begin{figure}
  \begin{center}
    \FigureFile(80mm,110mm){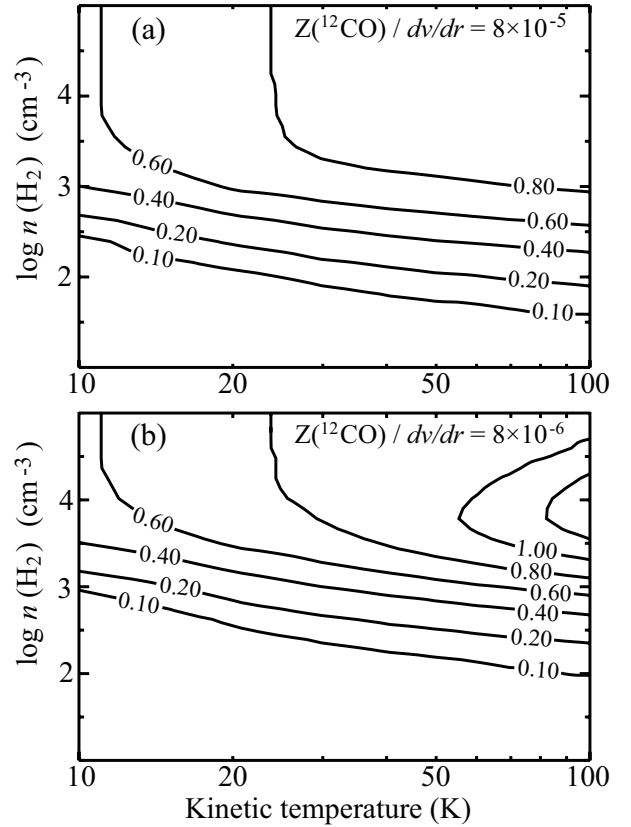}
  \end{center}
\caption{
Curves of constant $R_{3-2/1-0}$ as functions of kinetic temperature and density of molecular gas.
(a)
CO fractional abundance per unit velocity gradient Z($^{12}$CO)/($dv/dr$) was set to $8 \times 10^{-5}$,
where Z($^{12}$CO) is defined as [CO]/[H$_2$], and the unit of $dv/dr$ is km s$^{-1}$ pc$^{-1}$.
Note that no contours showing $R_{3-2/1-0}$ of 1 are seen.
(b)
Z($^{12}$CO)/($dv/dr$) was set to $8 \times 10^{-6}$.
}\label{fig:lvg}
\end{figure}

\subsection{Relation between the $R_{3-2/1-0}$ and SFE}

The observed radial variations of $R_{3-2/1-0}$ and the LVG calculation suggest
the fraction of the dense molecular gas at the nucleus is higher than that in the disk region in M~83.
Here we investigate the physical relationship between the molecular gas properties
traced by $R_{3-2/1-0}$ and the massive star formation. The first step is to quantify the
star formation rates (SFRs) and star formation efficiencies (SFEs) in M~83.

\subsubsection{Derivation of SFR and SFE}

An SFE is expressed using an SFR or a surface density of SFR ($\Sigma_{\rm SFR}$) and 
a surface mass density of molecular hydrogen ($\Sigma_{\rm H_2}$) as follows:

\begin{eqnarray}
\left[ \frac{\rm SFE}{\rm yr^{-1}} \right]= \left( \frac{\Sigma_{\rm SFR}}{M_{\odot}\,{\rm yr^{-1}\,pc^{-2}}} \right) {\displaystyle \biggl/} \left( \frac{\Sigma_{\rm H_2}}{M_{\odot}\,{\rm pc^{-2}}} \right) 
\end{eqnarray}

To calculate $\Sigma_{\rm SFR}$, we used a 6 cm radio continuum map obtained with the VLA
(\cite{ondrechen1985}), as shown in figure 5.
We derived a radial distribution of the 6 cm radio luminosity.
\citet{condon1992} relates the nonthermal luminosity $L_{\rm N}$ and SFR as follows:

\begin{eqnarray}
\left[ \frac{{\rm SFR}\,(M \geq 5 M_{\odot})}{M_{\odot}\,{\rm yr^{-1}}} \right] \sim 1.9 \times 10^{-22} \left( \frac{L_{\rm N}}{\rm W\,Hz^{-1}} \right) \nonumber \\ 
\times \left( \frac{\nu}{\rm GHz} \right)^{\alpha}
\end{eqnarray}
where $\alpha \sim 0.8$ is a nonthermal spectral index and $\nu$ is the observed frequency.
By assuming that all the observed 6 cm radio continuum flux of M~83 originate from
nonthermal emission due to supernova remnants, we converted the radial distribution of
6 cm radio luminosity into that of $\Sigma_{\rm SFR}$ by employing equation (2).
This is shown in figure 10a.
It is similar to those of CO($J=3-2$) and CO($J=1-0$) integrated intensities; there is a strong peak
at the center. In addition, we can see a secondary peak at the bar-end ($r$ $\sim$ 2.4 kpc).
The azimuthally averaged SFRs are about $1 \times 10^{-7}$ $M_{\odot}\,{\rm yr^{-1}\,pc^{-2}}$ in the disk region
and about $3 \times 10^{-6}$ $M_{\odot}\,{\rm\,yr^{-1}\,pc^{-2}}$ at the nucleus of M~83,
i.e., $\Sigma_{\rm SFR}$ at the nucleus where a moderate starburst occurs is enhanced by a factor of
$\sim 30$ as compared with that in the disk region.

$\Sigma_{\rm H_2}$ is estimated from the CO($J=1-0$) intensity
using a $N_{\rm H_2}$ / $I_{\rm CO}$ conversion factor $X_{\rm CO}$.
Here, we adopted a Galactic $X_{\rm CO}$ value of
$1.8 \times 10^{20}$ ${\rm cm}^{-2}$ (K km s$^{-1}$)$^{-1}$ \citep{dame2001}.
Note that the effect of the constant $X_{\rm CO}$ will be discussed at the end of this subsection.
$\Sigma_{\rm H_2}$ is calculated as

\begin{eqnarray}
\left[ \frac{\Sigma_{\rm H_2}}{M_{\odot}\,{\rm pc^{-2}}} \right] &=& 2.89 \times {\rm cos}(i) \left( \frac{I_{\rm CO}}{{\rm K\,km\,s^{-1}}} \right) \nonumber \\
&& \times \left\{ \frac{X_{\rm CO}}{1.8 \times 10^{20}\,{\rm cm}^{-2}\,({\rm K\,km\,s^{-1}})^{-1}} \right\} 
\end{eqnarray}
where $I_{\rm CO}$ is the CO($J=1-0$) intensity and $i$ is the inclination of 
the galaxy ($i$ = 24$^{\circ}$ for M~83, \cite{comte1981}).

Figure 10b shows the resultant azimuthally averaged SFEs as a function of
the galactocentric radius using equation (1).
In contrast with the radial profiles of CO and $\Sigma_{\rm SFR}$,
the radial profile of SFE in figure 10b shows a remarkable difference;
there is no enhancement of SFE at the bar-end ($r$ $\sim$ 2.4 kpc).
In the nuclear region, SFE is $8.9 \times 10^{-9}$ ${\rm yr}^{-1}$,
whereas the average SFE is $2.2 \times 10^{-9}$ ${\rm yr}^{-1}$ in the disk region,
i.e., four times lower than that in the nuclear region.
Considering the fact that these derived SFR and SFE values are limited to contributions from 
massive stars heavier than 5 $M_{\odot}$, the total (i.e., including low-mass stars)
SFE at the nucleus would be of an order of a few $\times$ $10^{-8}$ yr$^{-1}$  (see \cite{condon2002}),
which is in fact within the range of circumnuclear starbursts \citep{kennicutt1998a}.
The SFEs in the disk region of M~83 are comparable to or slightly higher than those of
nearby gas-rich spiral galaxies (e.g., \cite{wong2002}).

%% fig. 10
\begin{figure}
  \begin{center}
    \FigureFile(80mm,104mm){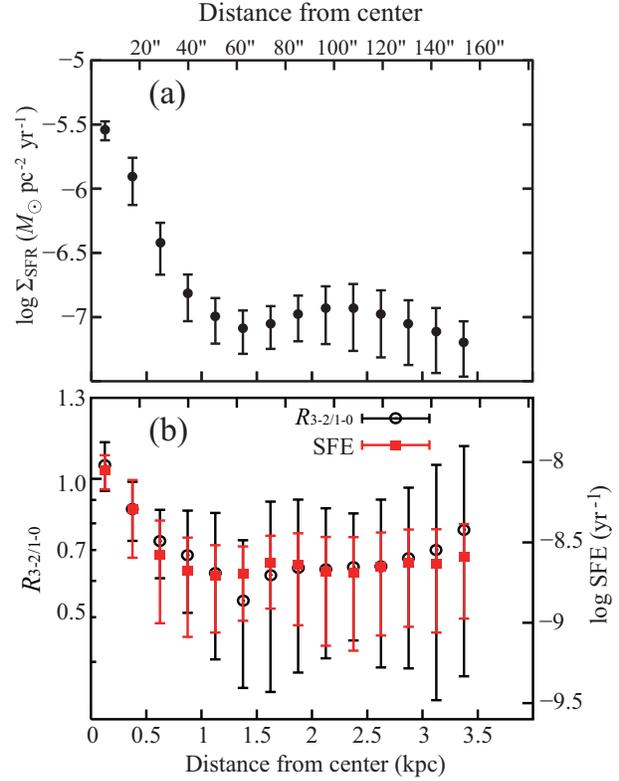}
  \end{center}
\caption{
(a)
The azimuthally averaged SFR as a function of the galactocentric radius of M~83.
The profile of SFRs show strong peaks in the central region and secondary peaks at the bar-end ($r \sim 2.4$ kpc)
as well as CO intensities shown in figure 6a.
(b)
The azimuthally averaged SFE and $R_{3-2/1-0}$ as a function of the galactocentric radius of M~83.
Both profiles show a strong peak at the center, whereas no significant enhancement is visible
at the bar-end, despite the fact that there is a secondary peak in the CO intensities and SFRs.
}\label{fig:sfrsfe}
\end{figure}

Note that we adopt a constant $X_{\rm CO}$ value over the observed region,
but our conclusion that there is an excess of SFE at the center of M~83 is not weakened
even after considering a possible variation of $X_{\rm CO}$ within M~83. 
In the central regions of various galaxies, $X_{\rm CO}$ factors tend to show smaller values
than those in the disk region by a factor of 2--3 or more 
(e.g., \cite{nakai1995}, \cite{regan2000}), including the Galactic
Center (e.g., \cite{oka1998}, \cite{dahmen1998}).
A smaller $X_{\rm CO}$ yields a smaller gas mass in the central region,
providing higher SFE than the present one in figure 10b.

\subsubsection{Comparison of $R_{3-2/1-0}$ with SFE}

Now we can compare the radial profile of $R_{3-2/1-0}$ with that of the SFE; figure 10b shows the results.
Here, we find a remarkable similarity in the radial profiles of $R_{3-2/1-0}$ and SFEs.
Both of them show a strong peak at the center, whereas no significant enhancement
can be seen at the bar-end, despite the fact that there is a secondary peak in the CO($J=1-0$) intensities,
corresponding to total mass of molecular gas, and the SFRs (figure 7a and figure 10a).
Further, figure 11 shows a correlation between $R_{3-2/1-0}$ and the SFEs.
These suggest that the physical state of molecular gas is intimately linked to
the property of massive star formation.

%% fig. 11
\begin{figure}
  \begin{center}
    \FigureFile(80mm,77mm){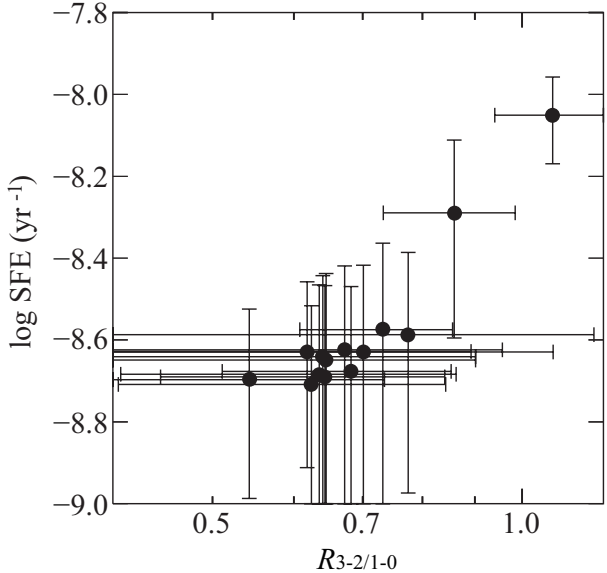}
  \end{center}
\caption{
The plot of the $R_{3-2/1-0}$ vs. SFE in each annuli of M~83.
A good correlation can be seen between $R_{3-2/1-0}$ and SFEs.
}\label{fig:ratiosfe}
\end{figure}

Let us compare the properties of star formation at the center and those at the bar-end.
At the bar-end, the amounts of molecular gas and massive stars are enhanced compared with other disk region;
however, there is no excess of SFE.
This means that the enhancement in $\Sigma_{\rm SFR}$ at the bar-end is
{\it solely} due to an increase in the amount of matter (i.e., molecular gas).
In contrast, we see a striking enhancement in the efficiency of star formation,
in addition to the elevated molecular gas mass and the rate of massive star formation
in the nuclear starburst region.
This means that the presence of nuclear starburst (i.e., the striking excess of star formation)
is not only due to the enhancement in gas mass but also due to the enhancement
in the {\it efficiency} of star formation. In other words, 
{\it a simple summation of star forming regions at the bar-end (and the disk region)
cannot reproduce star forming regions in the nuclear region of M~83}.

What causes this difference between the star formation properties in the nuclear starburst and disk star formation in M~83?
It is intriguing to see evidence that both $R_{3-2/1-0}$ and SFE follow the same trend (figure 10b and figure 11).
This may lead us to speculate that the spatial variation of a dense gas fraction governs the spatial variation of SFE in M~83;
there might be a threshold fraction, and once a dense gas fraction exceed the critical value, the SFE then increase there.

Of course, there is an opposite possibility; an excess of $R_{3-2/1-0}$ could be a {\it consequence},
not a cause, of the elevated efficiency of star formation at the center.
The strong UV radiation and successive supernovae within a starburst region could drastically
change the physical properties of the molecular gas.
Further multi-line/transition observations and their modeling using LVG and PDR codes will be required to
understand the origin and interpretation of the observed $R_{3-2/1-0}$ and SFE correlation in M~83.

\section{Summary}

We have performed the CO($J=3-2$) emission observations with the ASTE toward
the nearby barred spiral galaxy M~83.
The size of CO($J=3-2$) map is about 5$^{\prime}$ $\times$ 5$^{\prime}$ ($6.6 \times 6.6$ kpc),
including the entire inner disk.
In this paper, we focused on the radial profiles of molecular gas properties
and their relation to star formation. The summary of this work is as follows.

\begin{enumerate}
\item Significant emission was detected at the nucleus, bar, and inner spiral arms.
We successfully resolved these structures at a resolution of 22$^{\prime \prime}$.
No significant emission lines were detected in the inter-arm region.
Our CO($J=3-2$) maps spatially correlated well with 6 cm radio continuum map,
corresponding to the distribution of recent star formation.

\item The luminosity of CO($J=3-2$) $L^{\prime}_{\rm CO(3-2)}$
is $9.3 \times 10^7$ K km s$^{-1}$ pc$^2$ at the center ($r$ $<$ 0.5 kpc)
and $4.2 \times 10^8$ K km s$^{-1}$ pc$^2$ in the disk region (0.5 $<$ $r$ $<$ 3.5 kpc), respectively.
The global $L^{\prime}_{\rm CO(3-2)}$ was $5.1 \times 10^{8}$ K km s$^{-1}$ pc$^2$.
This means that $L^{\prime}_{\rm CO(3-2)}$ in the disk region significantly contributes
to the global $L^{\prime}_{\rm CO(3-2)}$.

\item The average $R_{3-2/1-0}$ at the 22$^{\prime \prime}$ resolution
was $\sim 1$ at the center of M~83 ($r$ $<$ 0.5 kpc).
The ratio drops to a constant value, 0.6--0.7, through the disk region (0.5 $<$ $r$ $<$ 3.5 kpc).

\item The radial profile of the SFE, determined from 6 cm radio continuum and the CO($J=1-0$) emission,
shows the same trend as $R_{3-2/1-0}$; i.e., the SFE shows a strong peak at the nucleus ($r$ $<$ 0.25 kpc),
whereas it drops to a constant value in the disk region (0.5 $<$ $r$ $<$ 3.5 kpc).

\item At the bar-end of M~83 ($r \sim 2.4$ kpc), the amounts of molecular gas and the massive stars are enhanced,
whereas there is no excess of $R_{3-2/1-0}$ and SFE in that region.
This means that the presence of nuclear starburst is not only due to the enhancement in the gas mass
but also due to the enhancement in the efficiency of star formation.
In other words, a simple summation of the star forming regions at the bar-end (and the disk region) 
cannot reproduce the nuclear starburst of M~83.
These results could suggest that the spatial variation of the dense gas fraction
traced by $R_{3-2/1-0}$ governs the spatial variation of SFE in M~83.
\end{enumerate} 

\vspace{0.5cm}

We would like to acknowledge the referee for his invaluable comments.
We thank the members of the ASTE team for the operation and
ceaseless efforts to improve the ASTE.
We are grateful to F. Comer{\'o}n for sending us the V-band image of M~83 obtained with VLT.
Observations with ASTE were (in part) carried out 
remotely from Japan by using NTT's GEMnet2 and its partner 
R\&E (Research and Education) networks,
which are based on AccessNova 
collaboration of University of Chile,
NTT Laboratories, and National Astronomical 
Observatory of Japan.
This study was financially supported by MEXT Grant-in-Aid
for Scientific Research on Priority Areas No. 15071202.
K. M., T.O., and A. E. are financially supported by a Grant-in-Aid for JSPS Fellows.

%%%%%%%%%%%%%%%%%%%%%%%%%%%%%%%%%%%%%%%

%%%
% See the manual for the detail.
%%%

\end{document}